\documentclass[aps,prd,reprint,amsmath,amssymb,superscriptaddress,nofootinbib,nolongbibliography,floatfix]{revtex4-2}
\usepackage{graphicx}
\usepackage{dcolumn}
\usepackage[utf8]{inputenc}
\usepackage{xcolor}
\usepackage[unicode=true,colorlinks,linkcolor=blue,anchorcolor=blue,urlcolor=blue,citecolor=blue,breaklinks=true]{hyperref}

\newcommand{\ii}{\mathrm{i}}

\newcommand{\pararrow}{\mathord{\buildrel{\lower3pt\hbox{$\scriptscriptstyle\leftrightarrow$}}\over {\partial}}} 
\newcommand{\pararrowk}[1]{\mathord{\buildrel{\lower3pt\hbox{$\scriptscriptstyle\leftrightarrow$}}\over {\partial}\hspace*{-0.18em}{}^#1}\hspace*{-0.18em} \,} 
\newcommand{\mytrace}[1]{\langle #1 \rangle} 

\usepackage{physics}

\newcommand{\yd}{\Upsilon_1(3\,{}^3D_1)}
\newcommand{\y}{\Upsilon(10753)}

\newcommand{\qfnu}{\affiliation{College of Physics and Engineering, Qufu Normal University, Qufu 273165, China}}

\newcommand{\imp}{\affiliation{Institute of Modern Physics, Chinese Academy of Sciences, Lanzhou 730000, China}}

\newcommand{\snst}{\affiliation{School of Nuclear Science and Technology, University of Chinese Academy of Sciences, Beijing 101408, China}}

\newcommand{\scnt}{\affiliation{Southern Center for Nuclear-Science Theory (SCNT), Institute of Modern Physics, Chinese Academy of Sciences, Huizhou 516000, Guangdong,
		China}}

\begin{document}
	
	\title{Production of $X_b$ via radiative transition of $\Upsilon(10753)$}
	
	\author{Shi-Dong Liu}
	\author{Hao-Dong Cai}\qfnu
	\author{Zu-Xin Cai}\qfnu \imp
	\author{Hong-Shuo Gao}\qfnu
	\author{Gang Li}\email{gli@qfnu.edu.cn} \qfnu
	\author{Fan Wang}\qfnu
	\author{Ju-Jun Xie} \imp \snst \scnt
	
\begin{abstract}
We studied the radiative transitions between the $\Upsilon(10753)$, the $S$-$D$ mixed state of the $\Upsilon(4S)$ and $\Upsilon_1(3\,{}^3D_1)$, and the $X_b$, the heavy quark flavor symmetry counterpart of the $X(3782)$ in the bottomonium sector. The radiative transition was assumed to occur through the intermediate bottom mesons, including $P$-wave $B_1^{(\prime)}$ mesons as well as the $S$-wave $B^{(*)}$ ones. The consideration of the $B_1^{(\prime)}$ mesons leads to the couplings to be in $S$-wave, and hence enhances the contributions of the intermediate meson loops. The radiative decay width for the $\Upsilon(10753)\to\gamma X_b$ is predicted to be order of $10~\mathrm{keV}$, corresponding to a branching fraction of $10^{-4}$. Based on the theoretical results, we strongly suggest to search for the $X_b$ in the $e^+e^-\to\gamma X_b$ with $X_b\to\pi\pi\chi_{b1}$ near $\sqrt{s}=10.754~\mathrm{GeV}$, and it is hoped that the calculations here could be tested by the future Belle II experiments.
		
\end{abstract}

\date{\today}

\maketitle
	\section{Introduction}\label{sec:intro}
	
Studies of exotic states have received considerable attention since 2003 when Belle collaboration observed the $X(3872)$ state in the $\pi^+\pi^-J/\psi$ invariant mass spectrum~\cite{choi2003PRL91-262001}. The candidates for exotic states are usually referred to collectively as \textit{XYZ} states, representing a group of particles that cannot be easily explained by the traditional quark model. Understanding the behavior and characteristics of these \textit{XYZ} states is crucial for advancing our knowledge of the strong force and its role in the structure of matter. With the development of the experimental techniques, a fair amount of \textit{XYZ} states have been directly observed, especially in the charmonium sector, for instance, the well-studied $X(3872)$~\cite{choi2003PRL91-262001,aubert2005PRD71-071103,abazov2004PRL93-162002,aaltonen2009PRL103-152001,chatrchyan2013J04-154,aaij2013PRL110-222001} and $Z_c(3900)$~\cite{ablikim2013PRL110-252001,liu2013PRL110-252002,xiao2013PLB727-366,abazov2018PRD98-052010}. However, in the bottomonium sector, there are only two exotic states, named as $Z_b(10610)$ and $Z_b(10650)$~\cite{bondar2012PRL108-122001a,adachi2012a[x-,krokovny2013PRD88-052016,garmash2015PRD91-072003}. More details on theoretical and experimental studies of the \textit{XYZ} states can be found in the reviews~\cite{brambilla2020PR873-1,guo2018RMP90-015004,lebed2017PPNP93-143,kalashnikova2019PU62-568,chen2016PR639-1,meng2023PR1019-2266,baru2017J06-158}.

The $X(3872)$ is the most well-studied exotic state either from the experimental or theoretical points of view, so that its properties are widely used as inputs to predict new hadronic states in the heavy quark sector. It has been well known that the $X(3872)$ mass is $(3871.65\pm 0.06)~\mathrm{MeV}$, extraordinarily close to the $D^{*0}\bar{D}^0$ threshold ($3871.69~\mathrm{MeV}$), and its quantum numbers $J^{PC}=1^{++}$~\cite{workman2022P2022-083C01}. Employing the heavy quark flavor symmetry of the $c$ and $b$ quarks, it's natural to expect that an analogous state with a mass near the $B^{*0}\bar{B}^0$ threshold ($10.604~\mathrm{GeV}$) exists in the bottomonium sector, with the same quantum numbers as the $X(3872)$ or other commonality. Such state is usually named for short as $X_b$~\cite{hou2006PRD74-017504}. The possible value of the $X_b$ mass has been calculated in the framework of tetraquark model~\cite{ebert2006PLB634-214,ali2010PLB684-28,matheus2007PRD75-014005} and using the mesonic molecule interpretation~\cite{tornqvist1994ZPC61-525,guo2013PRD88-054007}, which lies between $10.5$ and $10.7~\mathrm{GeV}$.

By following the successful way to understanding the $X(3872)$, the productions and decays of the $X_b$ have been investigated extensively. The partial decay width for the process $X_b \to \omega \Upsilon(1S)$, for instance, was theoretically predicted to be tens of keVs~\cite{li2015PRD91-034020}, under the interpretation that the $X_b$ is a $B^{*}\bar{B}$ bound state. However, searching for the $X_b$ in various experiments appears to be fruitless~\cite{brambilla2020PR873-1} (and references therein). Clear signal of the $X_b$ was not observed in the $\pi^+\pi^-\Upsilon(1S)$~\cite{aad2015PLB740-199,chatrchyan2013PLB727-57} or $\pi^+\pi^-\pi^0\Upsilon(1S)$~\cite{he2014PRL113-142001,adachi2023PRL130-091902} invariant mass distribution, while the $X(3872)$ was clearly observed in the similar distributions of pions plus the $J/\psi$. Therefore, to search for the $X_b$, we need to find other possible channels, for instance the $X_b\to\pi\pi\chi_{bJ}$, of which the partial width has been predicted to reach tens of keVs~\cite{jia2023a[x-}.

The production of the $X(3872)$ in the radiative decay of higher charmonium(-like) was firstly observed by the BESIII Collaboration in 2014~\cite{ablikim2014PRL112-092001}. This observation is consistent with the early theoretical prediction for the radiative transition process between the $1^{--}$ charmonium states and the $X(3872)$, where the $1^{--}$ should be a $D$-wave charmonium or a $D_1\bar{D}$ molecule (the $Y(4260)$, for instance) \cite{guo2013PLB725-127}. Thanks to the high mass of the $X_b$, its production via the radiative decays of higher bottomonia, especially the $1^{--}$ states, is greatly expected. However, the theoretical production ratio of the $X_b$ as a $B^{*}\bar{B}$ molecule in the processes $\Upsilon(5S,\, 6S)\to\gamma X_b$ is only of the order $10^{-6}$~\cite{wang2023EPJC83-186}, and thus the experimental observation seems difficult. Imitating the case of the $X(3872)$, the newly observed $\y$~\cite{mizuk2019J10-220,adachi2023PRL130-091902} is favored for the $X_b$ production since the $\y$ is likely to be $S$-$D$ mixed state of the $\Upsilon(4S)$ and $\yd$~\cite{li2021PRD104-034036,li2022PRD105-114041,bai2022PRD105-074007,liu2024PRD109-014039,adachi2023a[x-}. It should be pointed out that there are also other interpretations for the $\y$, such as the tetraquark state \cite{wang2019CPC43-123102,bicudo2021PRD103-074507,bicudo2023PRD107-094515,ali2020PLB802-135217} or hybrid bottomonium with excited gluonic degrees of freedom \cite{brambilla2020PR873-1,tarruscastella2021PRD104-034019}. More new experimental studies about the $\y$ can be found in Refs. \cite{adachi2024a[x-,jia2023CPL40-121301}.

In this work, we calculated the production of the $X_b$ in the radiative process $\y \to \gamma X_b$ using a nonrelativistic effective field theory. We regarded the $X_b$ as the $B^{*}\bar{B}$ molecule and the $\Upsilon(10753)$ as the $4S$-$3D$ mixed state. Moreover, the radiative transition was assumed to occur through the intermediate bottom meson loops, including the $P$-wave $B_1^{(\prime)}$ mesons as well as the $S$-wave $B^{(*)}$ ones. The loops including the $B_1^{(\prime)} $ mesons are enhanced due to the $S$-wave couplings. 

The rest of the paper is organized as follows. In Sec.~\ref{sec:formula}, we present the theoretical framework used in this work. Then in Sec.~\ref{sec:results} the numerical results are presented, and a brief summary is given in Sec.~\ref{sec:summary}.
	
\section{Theoretical Consideration}   \label{sec:formula}
	
Similar to the previous works as done in Refs.~\cite{li2021PRD104-034036,li2022PRD105-114041,bai2022PRD105-074007,liu2024PRD109-014039}, we interpret the $\y$ as a $4S$-$3D$ mixed state, then the wave function of the $\Upsilon(10753)$ is written as
	\begin{equation}\label{eq:wf10753}
		\tilde{\Upsilon}(10753) = \tilde{\Upsilon}(4S) \sin\theta  + \tilde{\Upsilon}_1(3\,{}^3D_1) \cos\theta\,,
	\end{equation}
	where $\theta$ is a mixing angle to describe the proportion of the partial waves. $\tilde{\Upsilon}(4S)$ and $\tilde{\Upsilon}_1(3\,{}^3D_1)$ describe the wave functions of the pure $\Upsilon(4S)$ and $\yd$ states, respectively. The $\Upsilon(10580)$ was usually regraded as the $4S$ state \cite{workman2022P2022-083C01}. When taking into account the $S$-$D$ mixing, the $\Upsilon(10580)$ 
could be interpreted as another $4S$-$3D$ mixture, and accordingly has the wave function
	\begin{equation}\label{eq:wfY10580}
		\tilde{\Upsilon}(10580) = \tilde{\Upsilon}(4S) \cos\theta  - \tilde{\Upsilon}(3\,{}^3D_1) \sin\theta\,.
	\end{equation}
The mixing angle $\theta$ can be obtained by fitting the well-measured dielectron decay width of the $\Upsilon(10580)$~\cite{badalian2010PAN73-138,li2021PRD104-034036,liu2024PRD109-014039}, which depends on its wave function and mass. The estimation yields the mixing angle $\theta=23.4^\circ\sim 36.1^\circ$. Such mixing angles indicate that the $\yd$ is the dominant component to form the $\Upsilon(10753)$ with proportion of about $65\%\sim 84\%$. In the following, in order to predict the decay width of the $\y\to \gamma X_b$, we shall adopt an angle of $\theta = 33^\circ$ for the $\Upsilon(4S)$ and $\yd$ mixing with masses of $10.612~\mathrm{GeV}$ and $10.675~\mathrm{GeV}$, respectively, which was predicted using the modified Godfrey-Isgur model~\cite{wang2018EPJC78-915}.

	
	\subsection{Intermediate Bottom Meson Loops}
	
\begin{figure}
	\centering
	\includegraphics[width=0.96\linewidth]{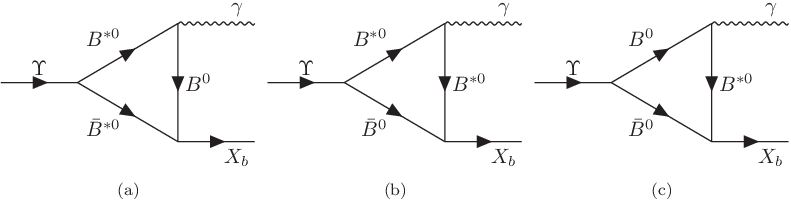}
	\caption{Triangle Feynman diagrams for the radiative process $\y \to \gamma X_b$ via the $S$-wave bottom meson loops. Each diagram has its corresponding charged one. The symbol $\Upsilon$ stands for the $\Upsilon(4S)$ and $\yd$.}
	\label{fig:feyndiagramsb}
\end{figure}

The bottomonia $\Upsilon(4S)$ and $\yd$ are both above the open-bottom threshold so that they are expected to dominantly decay into the bottom-antibottom meson pair, and then the pair could couple further to the final states by exchanging a proper bottom meson. This process is widely described by the triangle meson loop mechanism, which proves to be important in the decays and productions of many heavy quarkonia and exotic states. In the case of the radiative transition of the $\y$ to $X_b$, the loops made of the $S$-wave bottom mesons with the quantum numbers $s_l^P = 1/2^+$ of the light degrees of freedom are shown in Fig.~\ref{fig:feyndiagramsb}.

\begin{figure}
	\centering
	\includegraphics[width=0.86\linewidth]{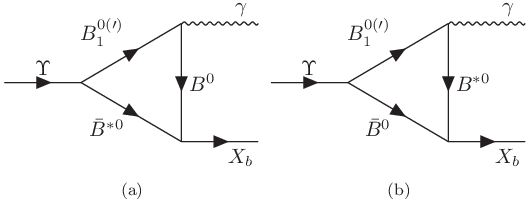}
	\caption{Feynman diagrams for the $\y \to \gamma X_b$ via the loops associated with the $P$-wave bottom mesons $B_1^{(\prime)}$.}
	\label{fig:feyndiagramsb1}
\end{figure}

In view of the quantum numbers $J^{PC} = 1^{--}$ for the initial bottomonia and final photon, they both couple to the $S$-wave $B^{(*)}$ mesons in a $P$ wave. The other coupling for the $X_b$ of $J^{PC}=1^{++}$ to the $S$-wave $B^{(*)}$ mesons occurs in an $S$ wave. When we replace the intermediate meson that connects the initial state $\Upsilon(4S)$ [$\yd$] and the photon by a $P$-wave bottom mesons $B_1^\prime $ with $s_l^P=1/2^+$ [$B_1$ with $s_l^P=3/2^+$], as shown in Fig.~\ref{fig:feyndiagramsb1}, all the couplings are then allowed to be in an $S$ wave. Near threshold, the $S$-wave contribution is usually more important than that from the $P$-wave. It should be pointed out that the coupled-channel effect due to the proximity of the $B^{(*)}\bar{B}^{(*)}$ threshold to the $\Upsilon(4S)$ and $\yd$ masses is highly suppressed by the $P$-wave couplings so that its absolute impact on the radiative transition $\y \to \gamma X_b$ is insignificant. Thus, in comparison with the loops in Fig.~\ref{fig:feyndiagramsb}, the contributions of the loops in Fig.~\ref{fig:feyndiagramsb1} to the radiative transition $\y\to \gamma X_b$ are likely to be more important because of the $S$-wave coupling enhancement, although the $\Upsilon(4S)$ and $\yd$ lie away from the $B_1^{(\prime)}\bar{B}^{(*)}$ threshold. This importance will be qualitatively analyzed in terms of the power counting and verified by the numerical calculations.

\subsection{Effective Lagrangians}

Similar to the case of the $X(3872)$ \cite{guo2015PLB742-394}, we also consider the $X_b$ as a pure molecule of the $B\bar{B}^* +\mathrm{c.c}$, of which the neutral and charged components are assumed to be of equal proportion, 
\begin{equation}
	\ket*{X_b} = \frac{1}{2} \left (\ket*{B^0\bar{B}^{*0}} + \ket*{\bar{B}^0B^{*0}} + \ket*{B^+B^{*-}} + \ket*{B^-B^{*+}} \right ) \,.
\end{equation}
The effective coupling of the $X_b$ to a pair of bottom and antibottom mesons is then given by
\begin{align}\label{eq:LagXb}
	\mathcal{L}_X &= \frac{1}{2}g_X^0 X_b^{\dagger i}(B^0\bar{B}^{*0i}+\bar{B}^0B^{*0i})\nonumber\\
	&+ \frac{1}{2} g_X^\mathrm{c} X_b^{\dagger i} (B^+B^{*-i} + B^-B^{*+i}) + \mathrm{H.c.}\,,
\end{align}
where the constants $g_X$'s describe the coupling strength of the $X_b$ to the bottom meson pairs. Here and later the symbols with (without) the dagger index represent the outgoing (incoming) fields relative to the coupling vertices.

The coupling, $g_X$, could be extracted from the binding energy ($\epsilon_X$) of the $X_b$ with respect to the mass threshold of its two components~\cite{guo2013PLB725-127}:
\begin{equation}\label{eq:couplegx}
	g_X = \left( \frac{16\pi}{\mu_B} \sqrt{\frac{2\epsilon_X}{\mu_B}} \right)^{1/2}\,,
\end{equation}
where $\mu_B = m_B m_{B^*}/(m_B+m_{B^*})$ and $\epsilon_B = m_B+m_{B^*} - m_{X_b}$ with the $m$'s being the masses of the particles indicated by the subscripts. Given the quite small mass difference of the neutral and charged bottom mesons, the $g_X^0$ and $g_X^\mathrm{c}$ are taken to be equal\footnote{According to the world average masses of the $B^0$ and $B^*$ \cite{workman2022P2022-083C01} and the predicted mass around $10.562~\mathrm{GeV}$ for the $X_b$~\cite{tornqvist1994ZPC61-525}, the relative difference of the $g_X^0$ and $g_X^\mathrm{c}$ do not exceed $0.5\%$.}. In addition, we adopt the binding energy $\epsilon_X$ from 2 to 100 MeV, similar to the values used in the recent work \cite{wang2023EPJC83-186,jia2023a[x-}.

Since we assume that the $\y$ is a mixture of the $\Upsilon(4S)$ and $\yd$, to calculate the decay width of the $\y\to\gamma X_b$ we should also know the interactions of the $S$- and $D$-wave bottomonia with the bottom mesons. Within the framework of the nonrelativistic effective field theory, the interactions of the $S$-wave bottomonia and the bottom-antibottom meson pair read~\cite{guo2011PRD83-034013,guo2010PRD82-034025,guo2013PLB725-127,wu2019PRD99-034022} 
\begin{align}\label{eq:LagS}
	\mathcal{L}_S &= \ii \frac{g_S}{2} \mytrace{\bar{H}_a^\dagger\vec{\sigma}\cdot\pararrow H_a^\dagger J}+ \frac{g_S^\prime}{2}\mytrace{  (\bar{H}_a^\dagger S_a^\dagger +\bar{S}_a^\dagger  H_a^\dagger) J } +\mathrm{H.c.}
\end{align}
Here $H_a = \vec{V}_a\cdot\vec{\sigma}+P_a$ and $S_a = \vec{V}_{1a}^\prime\cdot\vec{\sigma}+P_{0a}$ are the spin doublets formed by the $1/2^-$ and $1/2^+$ bottom mesons. 
In this work, $\vec{V}_a = (B^{*0},\, B^{*+})$ and $ P_a = (B^0,\,B^+)$ denoting the vector and pseudoscalar bottom mesons with $s_l^P=1/2^-$, respectively, while $V_{1a}^{\prime} = (B_1^{\prime 0},\,B_1^{\prime +})$ and $P_{0a} = (B_0,\,B_0^+)$ for the $s_l^P=1/2^+$ bottom mesons. Using the convention in Ref. \cite{guo2013PLB725-127}, the charge conjugated fields for the heavy bottom mesons are $\bar{H}_a = -\vec{\bar{V}}\cdot\vec{\sigma} + \bar{P}_a$ and $\bar{S}_a = - \vec{\bar{V}}_{1a}^\prime\cdot\vec{\sigma} + \bar{P}_{0a}$.
The $J=\vec{\Upsilon}\cdot\vec{\sigma} + \eta_b$ represents the spin doublet of the $S$-wave bottomonia $\Upsilon(nS)$ and $\eta_b(nS)$. Conventionally, $A\pararrow B \equiv A(\partial B) - (\partial A) B$, the $\vec{\sigma}$ stands for the Pauli matrices, and the subscript $a$ is the light flavor index.  After tracing operation in spinor space (indicated by $\mytrace{\cdots}$), the Lagrangian in Eq. \eqref{eq:LagS} is explicitly written as
\begin{align}\label{eq:LagSexpand}
	\mathcal{L}_S &= \ii g_S \Upsilon^i (\bar{V}_a^{\dagger j}\pararrowk{i}V_a^{\dagger j}-\bar{V}_a^{\dagger j}\pararrowk{j}V_a^{\dagger i} - \bar{V}_a^{\dagger i}\pararrowk{j}V_a^{\dagger j})\nonumber\\
	&- g_S \epsilon^{ijk} \Upsilon^i (\bar{V}_a^{\dagger k} \pararrowk{j} P_a^\dagger + \bar{P}_a^\dagger \pararrowk{j}V_a^{\dagger k}) +\ii g_S\Upsilon^i \bar{P}_a^\dagger \pararrowk{i}P_a^\dagger \nonumber\\
	&+ \epsilon^{ijk}g_S \eta_b \bar{V}_a^{\dagger i}\pararrowk{j}V_a^{\dagger k} +\ii g_S \eta_b (\bar{P}_a^\dagger \pararrowk{i}V_a^{\dagger i}-\bar{V}_a^{\dagger i}\pararrowk{i}P_a^\dagger) \nonumber\\
	&+g_S^\prime \eta_b (\bar{P}_{0a}^\dagger P_a^\dagger + \bar{P}_a^\dagger P_{0a}^\dagger )-g_S^\prime \eta_b (\bar{V}_{1a}^{\prime \dagger i}V_a^{\dagger i} + \bar{V}_a^{\dagger i} V_{1a}^{\prime\dagger i})  \nonumber\\
	& + g_S^\prime \Upsilon^i (\bar{P}_a^\dagger V_{1a}^{\prime\dagger i} + \bar{P}_{0a}^\dagger V_a^{\dagger i} - \bar{V}_a^{\dagger i} P_{0a}^\dagger - \bar{V}_{1a}^{\prime\dagger i} P_a^\dagger)\nonumber\\
	&+\ii g_S^\prime \epsilon^{ijk} \Upsilon^i (\bar{V}_a^{\dagger k} V_{1a}^{\prime\dagger j} + \bar{V}_{1a}^{\prime\dagger k} V_a^{\dagger j})+\mathrm{H.c.},
\end{align}
where the coupling constants $g_S$ and $g_S^\prime$ will be determined later.

The interactions of the $D$-wave bottomonium $\yd$ with a pair of bottom and antibottom mesons are written as~\cite{guo2013PLB725-127}
\begin{align}\label{eq:LagD}
	\mathcal{L}_D &= \ii \frac{g_D}{2}\mytrace{\bar{H}_a^\dagger \sigma^i \pararrowk{j} H_a^{\dagger} J^{ij}}\nonumber\\
	&+ \frac{g_D^\prime}{2} \mytrace{(\bar{T}_a^{\dagger j}\sigma^i H_a^\dagger - \bar{H}_a^{\dagger}\sigma^i T_a^{\dagger j})J^{ij}}+ \mathrm{H.c.}\,,
\end{align}
where $J^{ij}$ represents the field for the $D$-wave bottomonium $\yd$ in the two-component notation \cite{guo2013PLB725-127}
\begin{equation}
	J^{ij} = \frac{\sqrt{15}}{10} (\Upsilon_1^i\sigma^j + \Upsilon_1^j\sigma^i) - \frac{1}{\sqrt{15}}\delta^{ij}\vec{\Upsilon}_1\cdot\vec{\sigma}\,,
\end{equation}
and $T_a^i$ is the field for the $s_l^P = 3/2^+$ bottom mesons \cite{guo2013PLB725-127},
\begin{equation}
	T^i_a = \sqrt{\frac{2}{3}} V_{1a}^i + \ii \frac{1}{\sqrt{6}}\epsilon^{ijk} V_{1a}^j\sigma^k\,.
\end{equation}
It should be pointed out that for the $s_l^P = 3/2^+$ mesons with the total angular-momentum of 2 are not considered in this work. Hence, $\vec{V}_{1a} = (B_1^0,\,B_1^+)$. The antibottom meson field $\bar{T}_{1a}^i$ is described as
\begin{equation}
	\bar{T}^i_a = \sqrt{\frac{2}{3}} \bar{V}_{1a}^i - \ii \frac{1}{\sqrt{6}}\epsilon^{ijk} \bar{V}_{1a}^j\sigma^k\,.
\end{equation}
The tracing evaluation yields the Lagrangian for the $\yd$,
\begin{align}\label{eq:LagDexpand}
	\mathcal{L}_D &= \ii g_D\frac{\sqrt{15}}{3}\Upsilon_1^i \bar{P}_a^\dagger \pararrowk{i}P_a^\dagger \nonumber\\
	&+  g_D\frac{\sqrt{15}}{6} \epsilon^{ijk} \Upsilon_1^i(\bar{P}_a^\dagger \pararrowk{j}V_a^{\dagger k} + \bar{V}_a^{\dagger k}\pararrowk{j}P_a^\dagger)\nonumber\\
	& +  \ii\frac{ g_D}{2\sqrt{15}}\Upsilon_1^i (4 \bar{V}_a^{\dagger j}\pararrowk{i}V_a^{\dagger j} - \bar{V}_a^{\dagger j}\pararrowk{j}V_a^{\dagger i} - \bar{V}_a^{\dagger i}\pararrowk{j}V_a^{\dagger j})\nonumber\\
	&+ g_D^\prime \frac{\sqrt{10}}{2} \Upsilon_1^i (\bar{V}_{1a}^{\dagger i} P_a^\dagger - \bar{P}_a^\dagger V_{1a}^{\dagger i})\nonumber\\
	& + \ii g_D^\prime \frac{\sqrt{10}}{4} \epsilon^{ijk}\Upsilon_1^i (\bar{V}_{1a}^{\dagger k} V_a^{\dagger j}+\bar{V}_a^{\dagger k} V_{1a}^{\dagger j})+\mathrm{H.c.}
\end{align}

The photonic coupling to the bottom mesons with $s_l^P = 1/2^-$ is written as~\cite{guo2013PLB725-127,hu2006PRD73-054003}
\begin{equation}\label{eq:Lagphoton}
	\mathcal{L}_{HH\gamma} = \frac{e\beta}{2}\mytrace{H_a^\dagger H_b\vec{\sigma}\cdot \vec{B}Q_{ab}} + \frac{eQ'}{2m_{Q'}}\mytrace{H_a^\dagger \vec{\sigma}\cdot\vec{B}H_a}\,,
\end{equation}
where $B^k = \epsilon^{ijk}\partial^i A^j$ is the magnetic field, $Q_{ab} = \mathrm{diag}(-1/3,\,2/3)$ denotes the charge matrix of the light $d$ and $u$ quarks, and $Q' = -1/3$ and $m_{Q'}$ stand for the $b$-quark charge and its mass, respectively. 

In addition, the radiative transition of the $1/2^+$ and $3/2^+$ bottom mesons to the $1/2^-$ ones is described by the following Lagrangian \cite{guo2013PLB725-127,wang2023EPJC83-186}
\begin{equation}\label{eq:Lagphotonp}
	\mathcal{L}_{S/TH\gamma} = - \frac{\ii e\tilde{\beta}}{2} \mytrace{H_a^\dagger S_b \vec{\sigma}\cdot \vec{E}Q_{ba}} +  \mytrace{T_a^iH_b^\dagger C_{ab}}E^i 
\end{equation}
where $E^i$ is the electric field. The $C_{ab}$ in the second term is a formalistic $2\times 2$ diagonal matrix in the form of $\mathrm{diag}(C^d,\,C^u)$, of which the elements describe the coupling strength. Explicitly, 

\begin{align}
	\mathcal{L}_{HH\gamma} &= \ii e \beta Q_{ab} \partial^i A^j (V_a^{\dagger i} V_b^j - V_a^{\dagger j}V_b^i) \nonumber\\
	&+ e\beta Q_{ab} \epsilon^{ijk} \partial^iA^j (P_a^\dagger V_b^k + V_a^{\dagger k} P_b)\nonumber\\
	& + \ii \frac{e Q^\prime}{m_{Q'}} \partial^i A^j(V_a^{\dagger j} V_a^i - V_a^{\dagger i}V_a^j) \nonumber\\
	&+ \frac{e Q^\prime}{m_{Q'}} \epsilon^{ijk} \partial^i A^j (V_a^{\dagger k}P_a+ P_a^\dagger V_a^k)
\end{align}
and
\begin{align}
	\mathcal{L}_{S/TH\gamma}&=e\tilde{\beta} Q_{ab}\epsilon^{ijk} \partial^0 A^i V_a^{\dagger j} V_{1b}^{\prime k} \nonumber\\
	&-\ii e \tilde{\beta} Q_{ab} \partial^0 A^i (V_a^{\dagger i}P_{0b} + P_a^\dagger V_{1b}^{\prime i})  \nonumber\\
	&+ \ii \sqrt{\frac{2}{3}}C_{ab}\epsilon^{ijk} \partial^0 A^i V_{1a}^jV_b^{\dagger k} \nonumber\\
	&+ 2\sqrt{\frac{2}{3}}C_{ab}\partial^0 A^i P_b^\dagger V_{1a}^i\,.
\end{align}

\section{Numerical results} \label{sec:results}

To proceed, we should evaluate the coupling constants $g$'s in Eqs.~\eqref{eq:LagS} and \eqref{eq:LagD}, the parameter $\beta$ in Eq. \eqref{eq:Lagphoton}, and, $\tilde{\beta}$ and $C_{ab}$ in Eq. \eqref{eq:Lagphotonp}. The way to evaluating the constants $g_S$ and $g_D$ is the same as that in our recent work~\cite{liu2024PRD109-014039} so that the details are not repeated. Notice that due to the factor $1/2$ in Eqs.~\eqref{eq:LagS} and \eqref{eq:LagD}, the values of $g_S$ and $g_D$ here are twice of those in Ref.~\cite{liu2024PRD109-014039}. Their values are summarized in Table~\ref{tab:globalgg}.

\begin{table}
	\caption{Coupling constants $g_S$ and $g_D$ (Units: $\mathrm{GeV^{-3/2}}$) we employed in the calculations. Their estimations are based on the theoretical and experimental data in Refs. \cite{wang2018EPJC78-915,workman2022P2022-083C01}.}
	\label{tab:globalgg}
	\begin{ruledtabular}
		\begin{tabular}{lccr}
			$g_S\,,g_D$	&$B\bar{B}$& $B\bar{B}^*+\mathrm{c.c.}$&$B^*\bar{B}^*$\\
			\colrule
			$\Upsilon(4S)$ & 0.776 & 0.776 & 0.776\\
			$\yd$ & 0.157 & 0.376 & 1.879\\
		\end{tabular}
	\end{ruledtabular}
\end{table}

However, the coupling constants $g_S'$ and $g_D'$ cannot be directly determined in terms of the way to the $g_S$ and $g_D$, since the threshold of the $B^{(*)}B_1^{(\prime)}$ exceeds the masses of the  $\Upsilon(4S)$ and $\yd$.
In order to give reasonable estimation of the $g_S'$ and $g_D'$, we then, considering the heavy quark symmetry, assume that the ratios $g_S'/g_S$ and $g_D'/g_D$ are heavy-flavor-independent, although the $g$'s are all heavy-flavor-dependent separately. Given the predictions for the $\psi(4S)\to D_1'\bar{D}$ \cite{gui2018PRD98-016010,wang2019PRD99-114003}, the value of $g_S'$ for the $\psi(4S)$ varies from 0.11 to 0.42 $\mathrm{GeV^{-1/2}}$, whereas the prediction for the $\psi(4S)\to D\bar{D}$ yields $g_S\approx 0.15~\mathrm{GeV^{-3/2}}$ for the $\psi(4S)$. As a result,  $g_S'/g_S$ is in the range $[0.73,\,2.8]~\mathrm{GeV}$. Similarly, in the case of the $\psi_1(3\,{}^3D_1)$, $g_D$ is estimated to be between $0.1$ and $0.3~\mathrm{GeV^{-3/2}}$, and $g_D'$ ranges from $0.2$ and $0.5~\mathrm{GeV^{-1/2}}$ according to the predictions in Refs.~\cite{gui2018PRD98-016010,wang2019PRD99-114003}. Considering the intermediate values, it gives the ratio $g_D'/g_D\approx 1.7~\mathrm{GeV}$.

In Refs.~\cite{choi2007PRD75-073016,zhu1997MPLA12-3027}, the radiative width for the $B^{*+}$ is $(0.40\pm 0.03)~\mathrm{keV}$ and for the $B^{*0}$ it is $(0.13\pm 0.03)~\mathrm{keV}$. Using the Lagrangian in Eq.~\eqref{eq:Lagphoton} together with the mass $m_b=4.18~\mathrm{GeV}$~\cite{workman2022P2022-083C01}, we get $\beta = 2.1~\mathrm{GeV^{-1}}$.
Likewise, based on the interactions between the $B_1^{(\prime)}$ and $B$ in Eq. \eqref{eq:Lagphotonp}
and the radiative widths of the $B_1^{(\prime)}\to\gamma B^{(*)} $ predicted in Ref. \cite{asghar2018EPJA54-127}, $\tilde{\beta}$ is estimated to be in the range$[1.5,\,2.0]~\mathrm{GeV^{-1}}$, and $C^u$ is between $0.22$ and $0.31~\mathrm{GeV^{-1}}$ for the $B_1^+\to\gamma B^{(*)+}$, while $C^d$ varies from $-0.17$ to $-0.12$ $\mathrm{GeV^{-1}}$ for the $B_1^0\to\gamma B^{(*)0}$. Notice that the minus sign is assigned for the later case, analogous to the radiative decay of $B^{*}$ meson \cite{choi2007PRD75-073016,zhu1997MPLA12-3027}.


\begin{figure}
	\centering
	\includegraphics[width=0.92\linewidth]{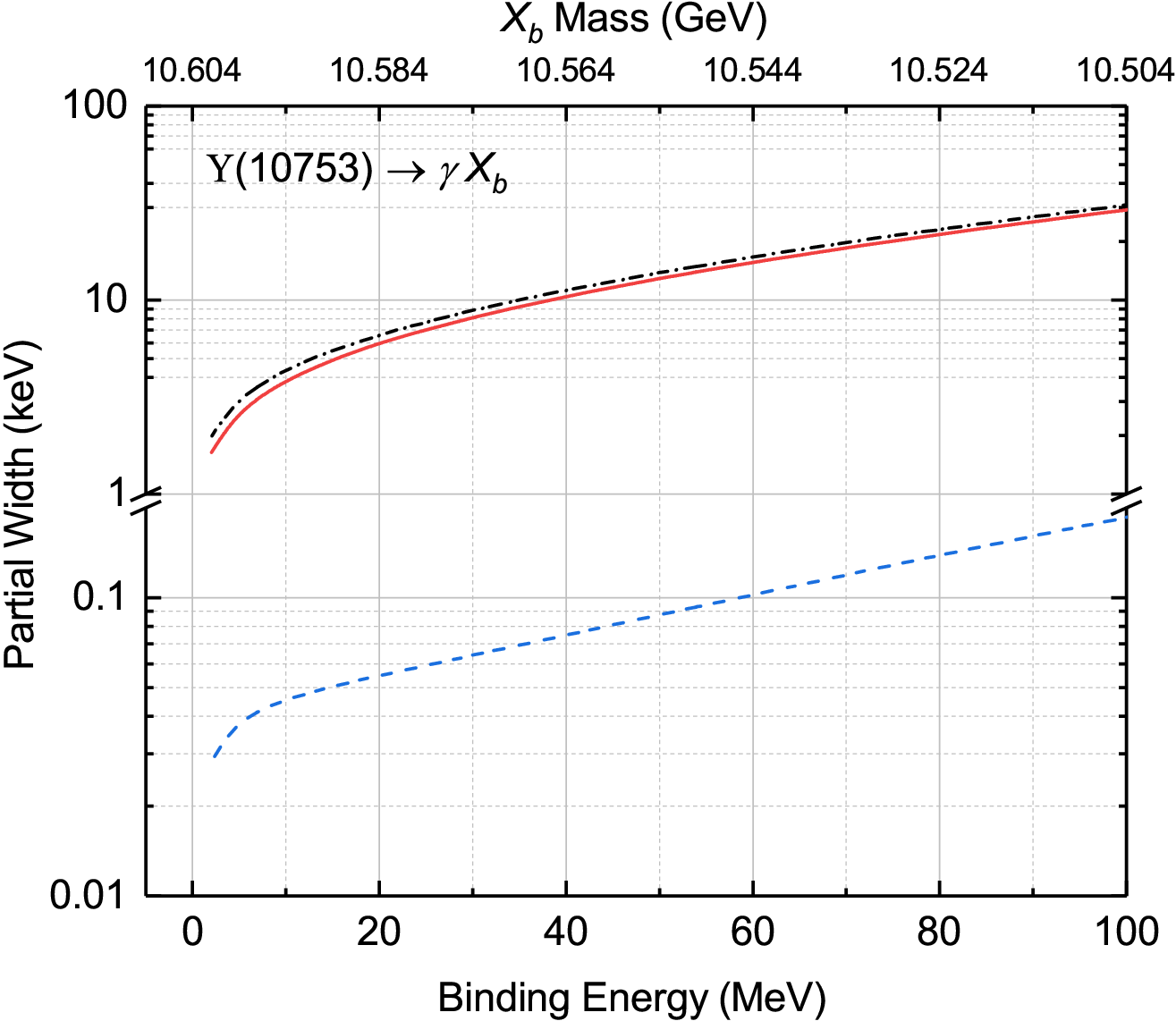}
	\caption{Partial decay width of the radiative transition $\y \to \gamma X_b$ for different binding energies of the $X_b$ relative to the $B^*\bar{B}$ threshold. In calculations, $g_S'$ is taken to be $1.5g_S$ and $g_D'$ is fixed to be $1.7g_D$. Moreover, $\tilde{\beta} =1.75~\mathrm{GeV^{-1}}$, $C^d = -0.15~\mathrm{GeV^{-1}}$, and $C^u=0.26~\mathrm{GeV^{-1}}$. The blue dashed and black dash-dotted lines describe, respectively, the contributions of the loops in Figs. \ref{fig:feyndiagramsb} and \ref{fig:feyndiagramsb1}, while the red solid line denotes all the loop contributions.}
	\label{fig:y10753toxb}
\end{figure}

Figure~\ref{fig:y10753toxb} exhibits the partial width for the radiative decay $\y\to \gamma X_b$ as a function of the $X_b$ binding energy from $2$ to $100~\mathrm{MeV}$, equivalently the mass between $10.504$ and $10.602~\mathrm{GeV}$ as indicated by the upper axis labels. It is seen that with increasing the binding energy, the radiative width increases. In particular, if the $X_b$ as a molecule assumes a binding energy of $50~\mathrm{MeV}$, corresponding to a mass of about $10.554~\mathrm{GeV}$, we predict the radiative width 
\begin{equation}
	\Gamma[\y\to\gamma X_b] \sim 13~\mathrm{keV}\,,
\end{equation}
which yields a branching fraction of order $10^{-4}$, two orders of magnitude larger than the processes $\Upsilon(5S,\, 6S)\to\gamma X_b$~\cite{wang2023EPJC83-186}.

When we look separately at the contributions from the loops in Figs.~\ref{fig:feyndiagramsb} and \ref{fig:feyndiagramsb1}, indicated by the blue dashed line and the black dash-dotted line, respectively, it is clearly shown that the radiative process $\y\to\gamma X_b$ is predominantly governed by the loops in Fig.~\ref{fig:feyndiagramsb1}. The dominance of the loops involving the $P$-wave $B_1^{(\prime)}$ mesons, compared to the loops composed entirely of the $S$-wave $B^{(*)}$ mesons, is consistent with the power counting: For the loops in Fig. \ref{fig:feyndiagramsb}, the initial vertex is in a $P$ wave and produces a momentum (see the Lagrangian in Eqs. \eqref{eq:LagSexpand} and \eqref{eq:LagDexpand}). This momentum has to be contracted with the external photon momentum $q$, and hence the initial vertex could be counted as $q$~\cite{guo2013PLB725-127}. As a result, within the nonrelativistic framework, the loop integral scales as~\cite{guo2013PLB725-127,wu2019PRD99-034022,wang2023EPJC83-186}
\begin{equation}\label{eq:loopsP}
	\frac{v^5}{(v^2)^3} q^2 = \frac{E_\gamma^2}{ v}\,.
\end{equation}
Here $v$ can be understood as the average of the intermediate bottom meson velocities. The velocity can be estimated by $\sqrt{2\abs*{m_1+m_2-M_{i(f)}}/(m_1+m_2)}$, where $m_1$ and $m_2$ are the masses of the bottom mesons related to the initial meson of mass $M_i$ or the final meson of mass $M_f$ \cite{guo2013PLB725-127}.

For the loops in Fig.~\ref{fig:feyndiagramsb1}, the initial vertex is in an $S$ wave thanks to the positive-parity bottom mesons $B_1^{(\prime)}$. In this case, the vertex is independent of the momentum (see Eqs. \eqref{eq:LagSexpand} and \eqref{eq:LagDexpand}). Therefore, such loop integral scales  as \cite{guo2013PLB725-127,wu2019PRD99-034022,wang2023EPJC83-186}
\begin{equation}\label{eq:loopsS}
	\frac{v^5}{(v^2)^3}q m_B = m_B\frac{E_\gamma}{v}\,,
\end{equation}
where $m_B$, the mass of the bottom meson, is introduced to balance the dimensions between Eqs. \eqref{eq:loopsP} and \eqref{eq:loopsS}. According to the estimations at the beginning of this section, the coupling constants for the diagrams in Figs. \ref{fig:feyndiagramsb} and \ref{fig:feyndiagramsb1} are nearly of the same order of magnitude. Thus the contributions from the loops in Fig. \ref{fig:feyndiagramsb1}, when compared to those in Fig. \ref{fig:feyndiagramsb1}, are enhanced by a factor of $m_B/E_\gamma\sim \order{30}$, agreeing with the numerical results shown in Fig. \ref{fig:y10753toxb}.

Theoretically, the $1/2^+$ bottom meson $B_1^\prime$ has a large width. The predictions in Ref. \cite{asghar2018EPJA54-127} show that $\Gamma_{B_1^\prime}$ is around 130 MeV, which is about twice times smaller than the width ($\sim 240~\mathrm{MeV}$) predicted in Ref. \cite{du2018PRD98-094018}. For the $3/2^+$, the width was predicted to be about 20 MeV, agreeing with the measured data between 27.5 and 31 MeV \cite{workman2022P2022-083C01}. In order to considering the width effect, especially for the $B_1^\prime$ mesons, we assume the mass spectrum to be described by the Breit-Wigner formula \cite{wang2023EPJC83-186,wu2019PRD99-034022,wu2021EPJC81-193},
\begin{equation}
	f(s,m,\Gamma) = \frac{1}{\pi} \frac{m\Gamma}{(s-m^2)^2+m^2\Gamma^2}\,.
\end{equation}
Here $s$ is the mass squared of the meson in question, $m$ is the central mass, and $\Gamma$ is the meson width. Then the amplitude is given by
\begin{equation}
	\mathcal{M} = \frac{1}{W}\int_{s_1}^{s_h} f(s,m_{B_1^{(\prime)}},\Gamma_{B_1^{(\prime)}}) \mathcal{M}(m_{B_1^{(\prime)}}\to \sqrt{s})\dd{s},
\end{equation}
where $\mathcal{M}(m_{B_1^{(\prime)}}\to \sqrt{s}) $ represents the amplitude expression without considering the $B_1^{(\prime)}$ width, but the $B_1^{(\prime)}$ mass is replaced by the square root of the integration variable, $\sqrt{s}$. Additionally, $W = \int_{s_1}^{s_h} f(s,m_{B_1^{(\prime)}},\Gamma) \dd{s}$ with $s_l = m_{B_1^{(\prime)}}^2$ and $s_h = (m_{B_1^{(\prime)}}+\Gamma_{B_1^{\prime}})^2$.

\begin{figure}
	\centering
	\includegraphics[width=0.92\linewidth]{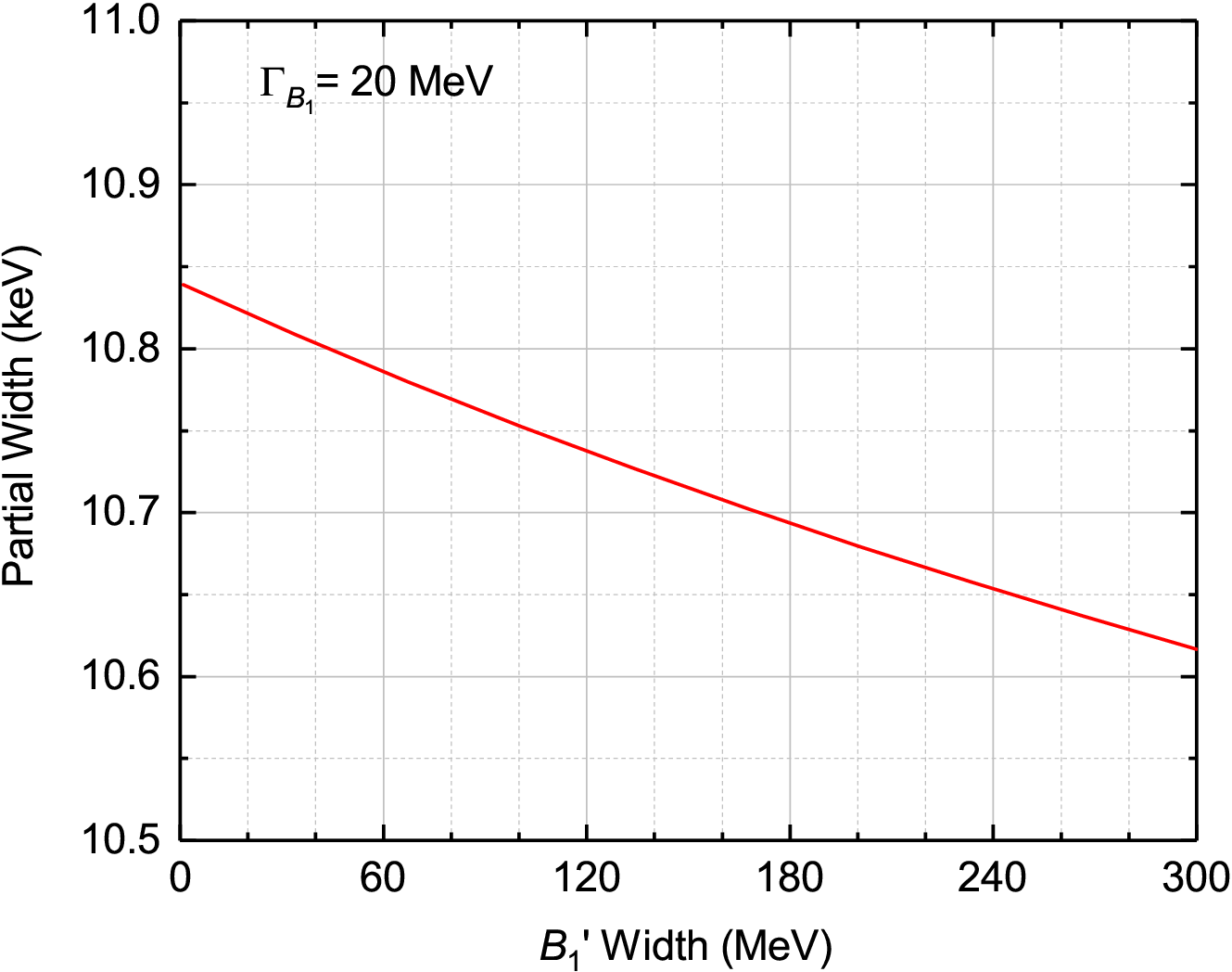}
	\caption{Radiative decay width for the process $\y\to \gamma X_b$ as a function of the $B_1^\prime$ width. The $B_1$ width is fixed to be $20~\mathrm{MeV}$. The couplings are the same as those in Fig. \ref{fig:y10753toxb}.}
	\label{fig:widthviswidth}
\end{figure}

In Fig. \ref{fig:widthviswidth}, the $B_1^\prime$ width dependence of the radiative decay width for the $\y\to\gamma X_b$ is shown. In the present calculations, the $B_1$ width is fixed to be $20~\mathrm{MeV}$ because of its smallness in comparison with that of the $B_1^\prime$. It is seen that the radiative decay width is slightly dependent on the $B_1^\prime$ width, decreasing less than $3\%$ when the $B_1^\prime$ width is increased to $300~\mathrm{MeV}$. It should be noted that the possible variation of the couplings resulting from the change of the $B_1^{(\prime)}$ width is not considered in our calculations, which might give rise to extra effect on the radiative decay width.

\section{Summary} \label{sec:summary}

In this work, we calculated the partial width for the radiative transition $\y\to\gamma X_b$, using a nonrelativistic effective field theory. In the calculations, we considered the $X_b$, the heavy quark flavor symmetry counterpart of the $X(3872)$ in the bottomonium sector, as a bound state of the $B^*\bar{B}+\mathrm{c.c}$, and the $\y$ as an $S$-$D$ mixed state of the $\Upsilon(4S)$ and $\yd$. Moreover, the radiative transition was assumed to occur through the intermediate bottom mesons, including the $P$-wave $B_1^{(\prime)}$ mesons as well as the $S$-wave $B^{(*)}$ ones.

It is found that the possible effect of the large width of the $B_1^\prime$ meson on the radiative decay width might be of minor importance, if the couplings do not change substantially with the $B_1^{(\prime)}$ width. Specially, our calculated results indicate that the radiative decay width is of order $10~\mathrm{keV}$ when the $X_b$ mass is around $10.56~\mathrm{GeV}$, corresponding a branching fraction of about $10^{-4}$. This bigness of the radiative width implies that searching for the $X_b$ via the process $\y\to\gamma X_b$ is promising. Recent experiments by Belle II Collaboration \cite{adachi2023PRL130-091902} did not find the $X_b$ in $e^+e^-\to\gamma X_b$ with $X_b\to\omega\Upsilon(1S)$ at $\sqrt{s}=10.745~\mathrm{GeV}$. However, given our recent study \cite{jia2023a[x-}, we suggest to hunt for the $X_b$ in the channel $e^+e^-\to\gamma X_b$ with $X_b\to\pi\pi\chi_{b1}$ near $\sqrt{s}=10.754~\mathrm{GeV}$.

\begin{acknowledgements}\label{sec:acknowledgements}

This work is partly supported by the National Natural Science Foundation of China under Grants No. 12105153, No. 12075133, No. 12047503, and No. 12075288, and by the Natural Science Foundation of Shandong Province under Grants No. ZR2021MA082 and No. ZR2022ZD26. It is also supported by Taishan Scholar Project of Shandong Province (Grant No.tsqn202103062), the Higher Educational Youth Innovation Science and Technology Program Shandong Province (Grant No. 2020KJJ004).	
\end{acknowledgements}

\bibliography{particlePhys.bib}
\end{document}